\documentclass[preprint2]{aastex}
\usepackage{natbib}
\shorttitle{AGNs at 90 GHz}
\shortauthors{Cotton et al.}

\begin{document}


\title{Observations of M87 and Hydra A at 90 GHz}


\author{W. D. Cotton, B. S. Mason}
\affil{National Radio Astronomy Observatory,
    520 Edgemont Road, Charlottesville, VA 22903-2475}
\email{bcotton@nrao.edu}

\author{S. Dicker, P. Korngut, M.J. Devlin, J. Aquirre}
\affil{University of Pennsylvania, 209 S. 33rd St., Philadelphia, PA 19104}

\author{D. Benford, H. Moseley, J. Staguhn\altaffilmark{1}}
\affil{NASA Goddard Space Flight Center, Greenbelt, MD 20771}


\author{K. Irwin}
\affil{National Institute of Standards and Technology, 325 Broadway, Boulder, CO 80303}

\and

\author{P. Ade}
\affil{School of Physics and Astronomy, Cardiff University, 5 The Parade, Cardiff, CF24 3AA, UK}


\altaffiltext{1}{also University of Maryland, College Park, MD
20742-2421}


\begin{abstract}
This paper presents new observations of the AGNs M87 and Hydra A at 90
GHz made with the MUSTANG bolometer array on the Green Bank Telescope
at 8.5$''$ resolution.  A spectral analysis is performed combining
this new data and archival VLA data on these objects at longer
wavelengths.  This analysis can detect variations in spectral index
and curvature expected from energy losses in the radiating particles.
M87 shows only weak evidence for steepening of the spectrum along the
jet suggesting either re-acceleration of the relativistic particles in
the jet or insufficient losses to affect the spectrum at 90 GHz.  The
jets in Hydra A show strong steepening as they move from the nucleus
suggesting unbalanced losses of the higher energy relativistic
particles.  The difference between these two sources may be accounted
for by the different lengths over which the jets are observable, 2 kpc
for M87 and 45 kpc for Hydra A.
\end{abstract}


\keywords{galaxies: jets, galaxies: active, radio continuum, galaxies: individual (M87, Hydra A), GBT}



\section{Introduction}

It has long been recognized that the radiative lifetimes of the
synchrotron emitting electrons in AGN jets and extended radio lobes
can be shorter than the time to transport them from the nucleus to
where they are observed  \citep{Felten68,Willis78}.  This
requires some resupply or re-acceleration of the radiating particles
well outside of the nucleus.  Observations of synchrotron jets visible
into the X-ray \citep{Biretta91} make this problem particularly acute.
For a synchrotron source of a given magnetic field strength, the
emission observed at a given frequency comes largely from electrons of
a given energy \citep{Pacholczyk70}.  Thus, the spectral shape of the
emission reflects the shape of the electron energy spectrum.  If the
predominant energy loss mechanism of these electrons is radiative
(synchrotron and inverse Compton scattering) then the losses are
proportional to the energy squared.  This depletes the number of
electrons above a corresponding energy resulting in a steepening of
the spectrum by 0.5 above a break frequency corresponding to the
highest energy surviving particles \citep{Pacholczyk70}.  If the
radiating electrons age in a constant magnetic field, the break
frequency will move to progressively lower frequencies.

Numerous attempts have been made to use this spectral signature to
determine the radiative ages of the particles, i.e. the time since
they were last accelerated to relativistic energies 
\citep{Carilli91,Blundell94,Feretti98,Murgia99}.
As was pointed out by \citet{Rudnick02}, this procedure has a number
of potential difficulties.
Most serious  is that the relationship of the frequency of the
spectral break with the cutoff in the electron energy spectrum depends
on a large power of the poorly known magnetic field
strength; \citet{Pacholczyk70} gives the relation among the break
frequency($\nu_B$), the magnetic field($B$) and time($t$) as:
$$ \nu_B\ =\ const\ B^{-3}t^{-2}\ \qquad\qquad {\rm 1)}.$$
Thus a lower magnetic field results in higher break
frequencies. 
While using spectra to determine particles ages is suspect, the
steepening of the spectrum along an AGN jet, especially at higher
frequencies shows the relative effects of particle losses and gains.

Most of the sources in which the signatures of spectral aging have
searched for have been observed over a relatively limited frequency range.
Since the effects of the electron energy losses are more pronounced at
higher frequencies, observations at high frequencies are valuable for
evaluating the relative effects of particle losses against any
re-acceleration or resupply.

In order to measure the variation of the observed spectrum along an
AGN jet, both good resolution and surface brightness sensitivity over
a wide range of frequencies is needed.  Observations with mm
interferometers have the resolution but generally not the surface
brightness sensitivity for comparisons with lower frequency
observations.  Since single dishes explicitly sample all spatial
frequencies out to the dish diameter, they have good surface
brightness sensitivity.  The large size of the Green Bank Telescope
(GBT) gives it good resolution as well as surface brightness
sensitivity.

In this paper we report on the observations of M87 and Hydra A at 90
GHz using the MUSTANG camera on 100 m the Green Bank Telescope \citep{jewellandprestage04}
which gives 8.5$''$ resolution and good surface brightness sensitivity.
These results are compared with lower frequency observations at
comparable resolutions using archival VLA data.
The results presented give the continuum spectra from 0.3 to 90 GHz;
this greatly increases the range of frequencies over which this analysis
can be performed.

\section {Observations} 
The radio sources in the AGNs M87 and Hydra A were observed on 2008
March 13 using the MUSTANG 3mm bolometer camera \citep{Mustang} on the
Green Bank Telescope (GBT).  The MUSTANG detector array consists of 64
TES bolometers viewing the sky through a filter with bandpass of 81 to
99 GHz.  Observations were made in the ``On--The--Fly'' (OTF) mode
using a variety of scanning patterns.  A cold field stop restricts the
portion of the GBT illuminated to 90 m.  The bolometer outputs are
recorded at 1 kHz but the data streams are averaged to 20 Hz for
analysis.  Focus and pointing calibration used J1256-0547 for M87 and
J0825+0309 for Hydra A.  Flux density calibration was based on
observations of planets made over the course of the season's
observations.

\section {MUSTANG Data Analysis}
Data analysis was carried out in the Obit package \footnote{http://www.cv.nrao.edu/$\sim$bcotton/Obit.html}
\citep{Obit08}.
Details of the data analysis will be given in a future publication but
a short description follows.

The principle difficulty in imaging the data from the MUSTANG array is
separating the portion of the signal in the time-stream due to the
astronomical object of interest  and the ``background'' portion due to
other sources, especially the atmosphere and the instrument itself.
Rapid scanning of the telescope over the target field helps separate
the signals in the time domain.
The redundancy resulting from the multiple observations of a given
pointing along different trajectories also helps separate the time
variable background signals from the constant astronomical signals.

The MUSTANG cryogenic system uses a pulse tube cooler with a
frequency of 1.4 Hz which imposes a strong modulation on the bolometer
outputs.
Prior to subsequent operations, the amplitude and phase of the
resultant signal were fitted and removed from the data.
Amplitude calibration used the response to a stable calibration signal
injected into all bolometers whose strength was calibrated against
observations of several planets.
Data weights were derived for each bolometer time-stream from the RMS
fluctuations during blank sky scans.

\subsection {Imaging\label{Imaging}}
Imaging used the general ``OTF'' technique of \citet{Mangum08} in
which the individual detector samples were multiplied by a continuous
``griding'' function and then re-sampled on a regular grid with $2''$
spacings.  For each grid cell, the weighted sum of the data times the
griding function and the sum of the weights times the griding function
were accumulated and the image value at that pixel was the set equal
to the ratio of the two.  The griding function used was an exponential
$\times$ sinc which helps suppress spatial frequencies larger than
those sampled by the dish and results in a resolution of 8.5$''$.
This process will produce a ``Dirty Image'', {\it i.e.}, an image of
the sky convolved with the telescope beam response.

\subsection {Background Estimation}
An iterative scheme of estimating the target brightness distribution,
subtracting a model from the time-stream data and (re)estimating the
backgrounds was employed.
In each iteration the shortest timescale used to estimate backgrounds
was reduced.

For the initial estimation of the backgrounds an empty sky was assumed
and backgrounds were estimated for timescales longer than 40 seconds.
This is longer than the crossing times of the target sources.
Following the initial background estimation, and for subsequent
iterations, the background subtracted time-stream data were imaged as
outlined in Section \ref{Imaging}.  A deconvolved image
(see \S~\ref{CLEAN}) was used to generate a sky model which was then
used with the assumed telescope point spread function (PSF) to
generate a residual time-stream data set from the initial calibrated
time-streams.  A refined background estimation was obtained from a low
pass filtering of the time-stream residuals.  This refined background
estimation was in turn used to derive a refined sky model.  Multiple
iterations of this process were used, gradually reducing the shortest
time scale in the background estimations to 1 second.  Most of the
backgrounds are common to all detectors.

\subsection {GBT Imaging Quality at 90 GHz}
At the epoch of these observations, uncorrected deformations of the
GBT surface caused a degradation of the telescope performance.
In particular, large scale errors in the surface result in an elevated
error beam in the neighborhood of the main beam.
While the near--in error beam is time variable in detail, general
features persisted over time.


If the point spread function (PSF) of the instrument is know and
stable, it is relatively straightforward to deconvolve the dirty
image and replace it with one in which the psf was well behaved.
The best estimate of the effective PSF of the telescope was obtained
from the averaged image of calibrator observations made during the
2008 Winter observing session.  A ``Dirty beam'' for the telescope was
derived by an azimuthal average of the average calibrator beam images
below the half power of the beam and an 8.5$''$ Gaussian above the
half power point.  This beam has an elevated error pattern near the
main beam and when used in a deconvolution, removes the scattered
power from around regions of bright emission. Since these data were
collected, a method of applying online corrections to the active
surface derived from out-of-focus beammaps has significantly reduced
the scattered-power problem.

\subsection {Deconvolution\label{CLEAN}}
The dirty image and dirty beam described in the previous sections
were deconvolved using a H\"ogbom CLEAN \citep{Hogbom03}.
This reduces the dirty image to a set of delta functions which can be
convolved by the GBT instrumental PSF to produce a sky model or
convolved with a Gaussian and restored to the CLEAN residuals to get the
final CLEAN image.
These CLEAN images restored with a 8.5$''$ FWHM Gaussian were used in
subsequent analysis.

One consequence of the way in which the dirty images were made is that
noise in the image pixels does not strictly have the same covariance
as the dirty beam.
This allows ``out of band'' noise which does not deconvolve and causes
the CLEAN to ``get stuck'', alternately digging and filling holes.
A filtering of the dirty image to remove all power corresponding to
spatial frequencies on smaller scales than allowed by the resolution
of the instrument greatly reduces this problem.


\begin{figure}
\centering
\includegraphics[height=3.25in,angle=-90]{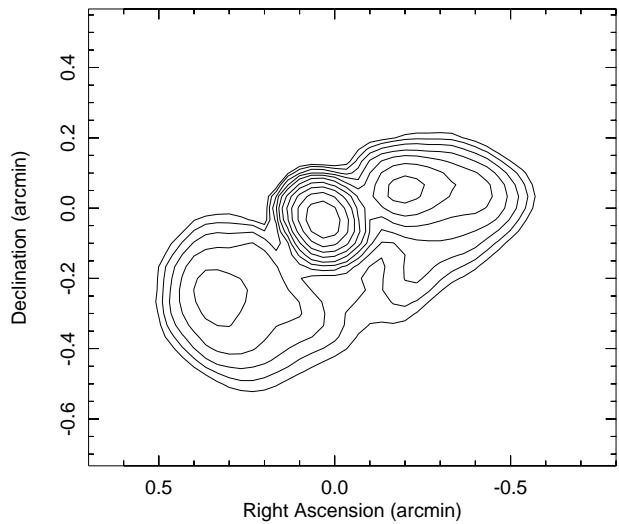}
\caption{ 
Cleaned Obit MUSTANG image of M87 at 8.5$''$ resolution.
Contours are powers of $\sqrt{2}$ times 0.1 Jy/beam.
}
\label{M87CleanFig}
\end{figure}

\begin{figure}
\centering
\includegraphics[height=3.0in,angle=-90]{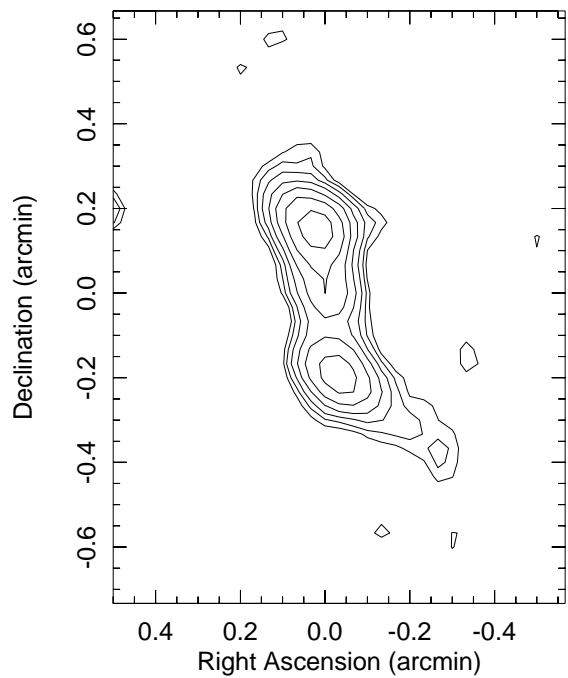}
\caption{ 
Cleaned Obit MUSTANG image of Hydra A at 8.5$''$ resolution.
Contours are powers of $\sqrt{2}$ times 0.025 Jy/beam.
}
\label{HydraACleanFig}
\end{figure}

The images of M87 and Hydra A derived in the Obit package are given in
Figures \ref{M87CleanFig} and \ref{HydraACleanFig}.

\section{VLA Data at Longer Wavelengths}
   In order to make a proper spectral analysis, images at comparable
resolution are needed over a range of frequencies.  
Fortunately, these bright radio galaxies have been well studied using
the VLA. 
Suitable resolution data were obtained from the VLA archives and
calibrated and imaged in the standard fashion using Obit.  
The imaging was initially at 8$''$ resolution using a combination of uv
ranges, tapering, and the Briggs robust parameter \citep{briggs95}.  
Subsequently, the images were convolved to 8.5$''$ for comparison with the
GBT data. 

\section{Spectral Decomposition}
The GBT and VLA images were all aligned to the same pixel grid and two sets
of spectral forms were fitted in each pixel by standard least squares
techniques.
These forms were a continiously curved spectrum with polynomials
in log(s) vs log($\nu$) using up to three terms; 1) the flux density at a
reference frequency, 2) the spectral index and 3) spectral curvature at
the same reference frequency.  
The other form was a broken power law where the spectral index at frequencies 
above the break steepened by 0.5.
All fitting used a reference frequency of 5 GHz.
For each source, a set of single point spectra were derived 
as well as images of the spectral index, spectral curvature and break frequency.

\begin{figure}
\centering
\centerline{
\includegraphics[height=2.5in,angle=-90]{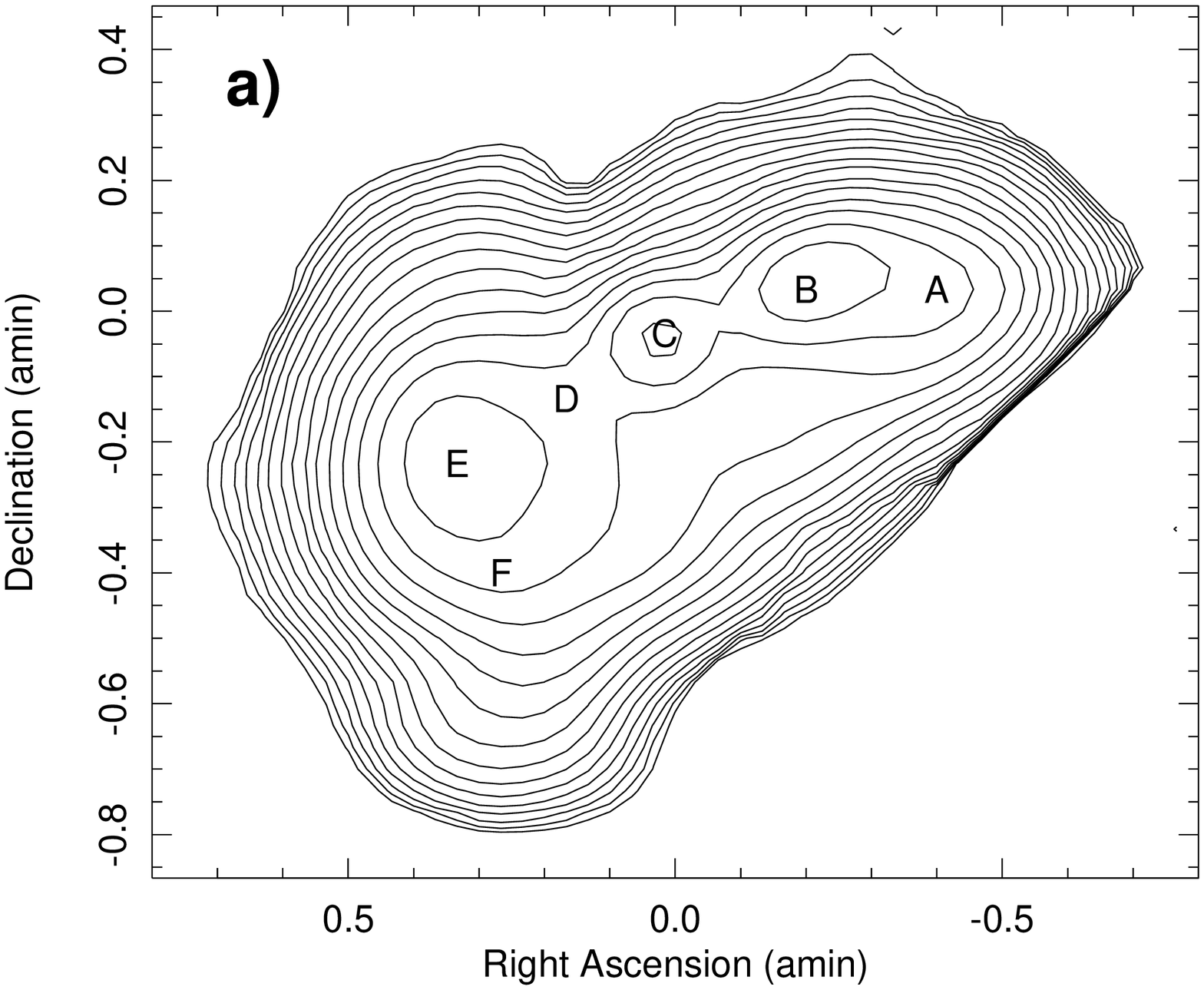}}
\centerline{
\includegraphics[height=2.4in]{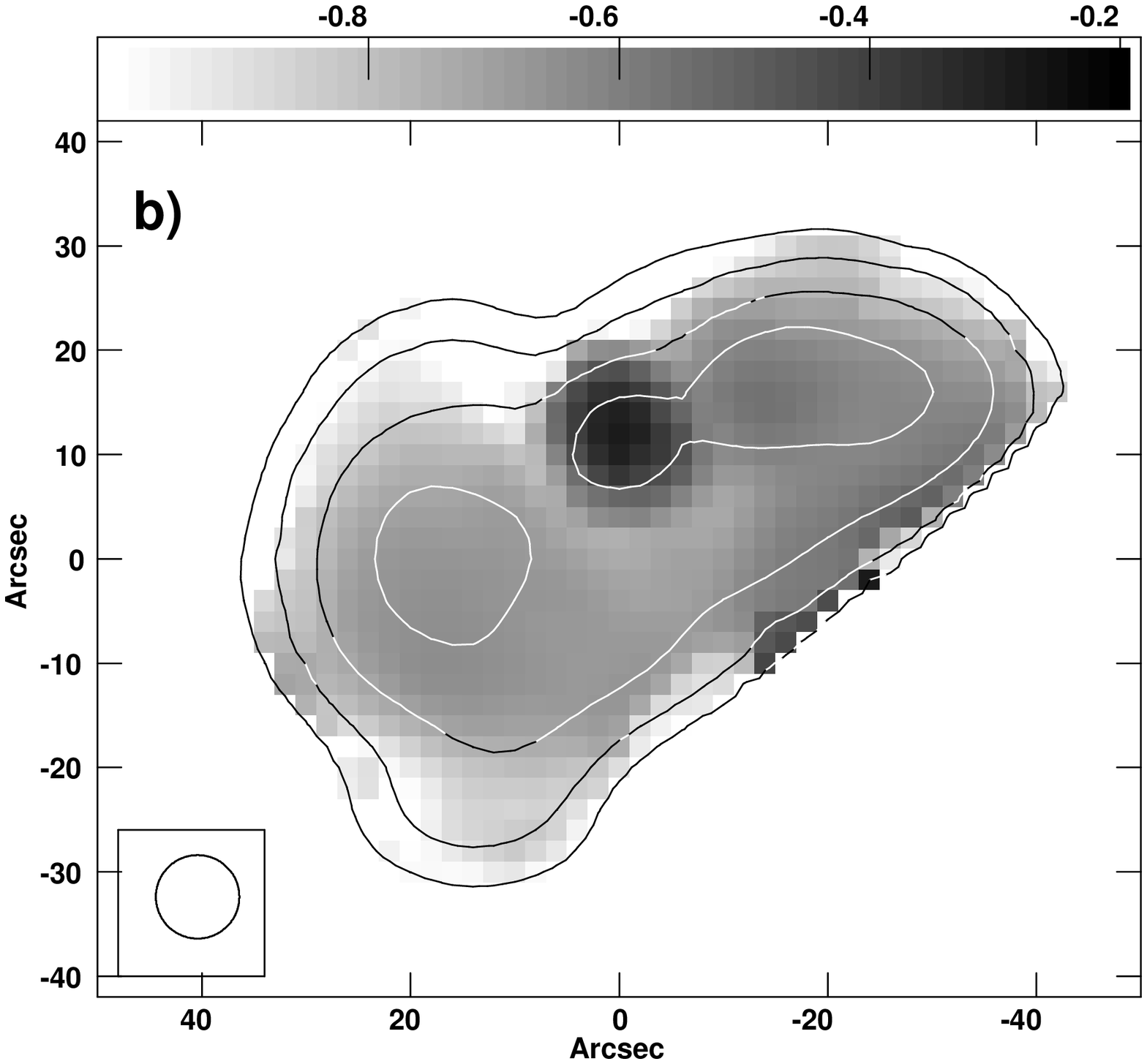}}
\centerline{
\includegraphics[height=2.4in]{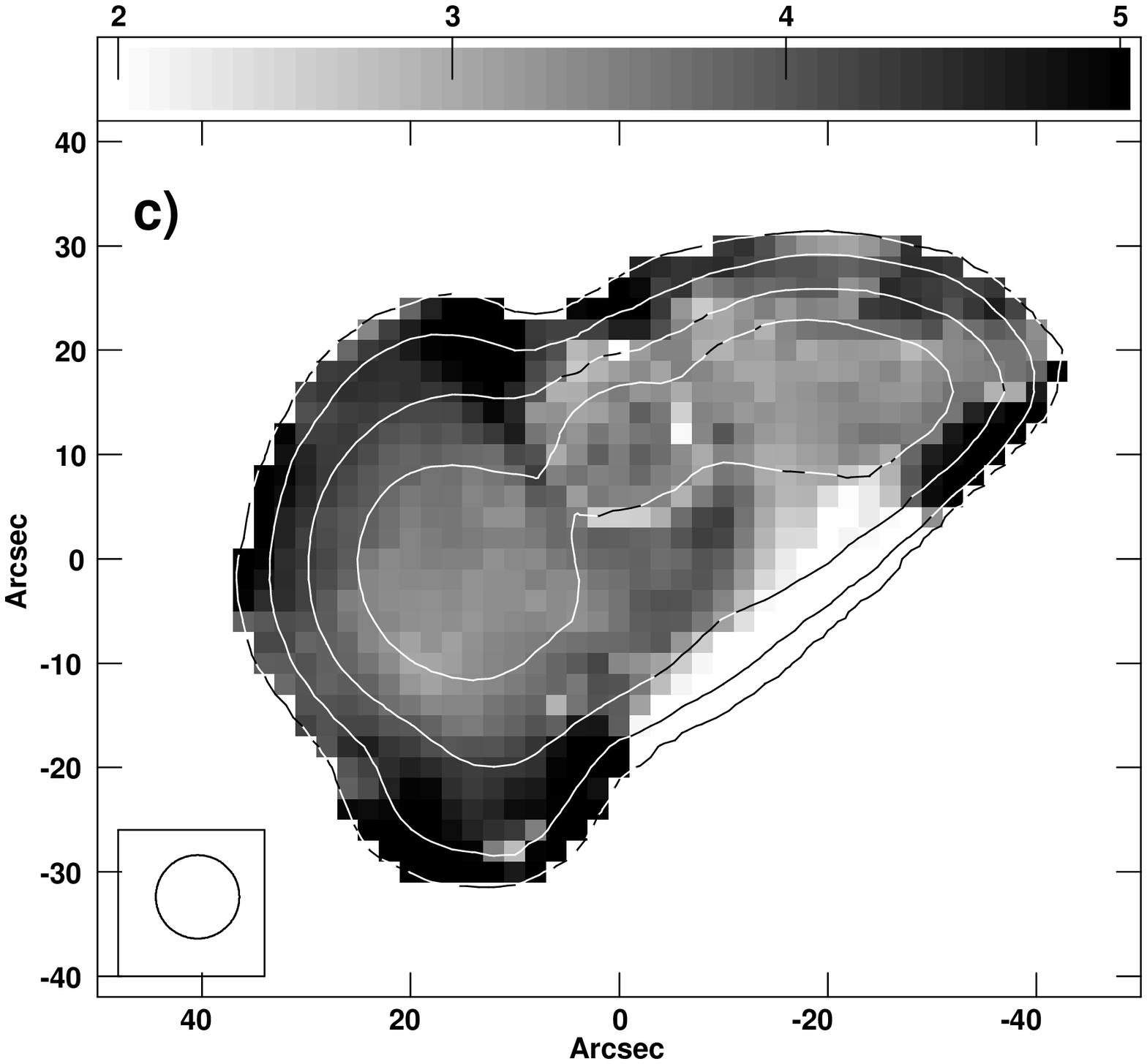}}
\caption{ 
{\bf a)} Fitted image of M87 at 5 GHz with indications of locations of point
spectra as letters.
Contours are powers of $\sqrt{2}$ times 25 mJy/beam.
{\bf b)} Grey-scale spectral index with contours of flux density at
100, 300, 
1000 and 3000 mJy/beam over-plotted.
The gray-scale is given by the scale-bar at the top.
The resolution is indicated by the circle in the lower left corner.
{\bf c)} as {\bf b)} but break frequency (GHz) shown in gray-scale.
}
\label{M87ContFig}
\end{figure}

\begin{table*}
\begin{center}
\caption{M87 Fitted Point Spectra at 5 GHz.\label{tblM87Pt}}
\begin{tabular}{crrrrrr}
\tableline\tableline
Posn. & Flux & Spectral& Curvature & $\chi^2_{\rm curve}$ & Break Freq. &$\chi^2_{\rm break}$\\
 & density (Jy) &  index &  &  & GHz \\
\tableline
A & 3.9 $\pm$ 0.2 & -0.62 $\pm$ 0.03 & -0.05 $\pm$ 0.01 & 1.5 & 3.3 $\pm$ 1.2 & 2.2 \\
B & 6.1 \phantom{$\pm$} 0.3 & -0.56 \phantom{$\pm$} 0.03 & -0.05 \phantom{$\pm$} 0.01 & 1.8 & 3.0 \phantom{$\pm$} 0.2 &  2.2 \\
C & 4.9 \phantom{$\pm$} 0.2 & -0.29 \phantom{$\pm$} 0.03 &  0.03 \phantom{$\pm$} 0.01 & 1.4 & 3.9 \phantom{$\pm$} 0.8 & 13.2 \\
D & 2.6 \phantom{$\pm$} 0.1 & -0.74 \phantom{$\pm$} 0.02 & -0.04 \phantom{$\pm$} 0.01 & 0.3 & 3.9 \phantom{$\pm$} 0.1 &  3.3 \\
E & 4.1 \phantom{$\pm$} 0.2 & -0.67 \phantom{$\pm$} 0.03 & -0.04 \phantom{$\pm$} 0.01 & 0.7 & 3.4 \phantom{$\pm$} 0.1 &  3.5 \\
F & 2.7 \phantom{$\pm$} 0.1 & -0.65 \phantom{$\pm$} 0.03 & -0.05 \phantom{$\pm$} 0.01 & 1.3 & 3.3 \phantom{$\pm$} 0.1 &  2.0 \\
\tableline
\end{tabular}
\end{center}
\end{table*}

\begin{figure*}
\centering
\centerline{
\includegraphics[height=3.0in,angle=-90]{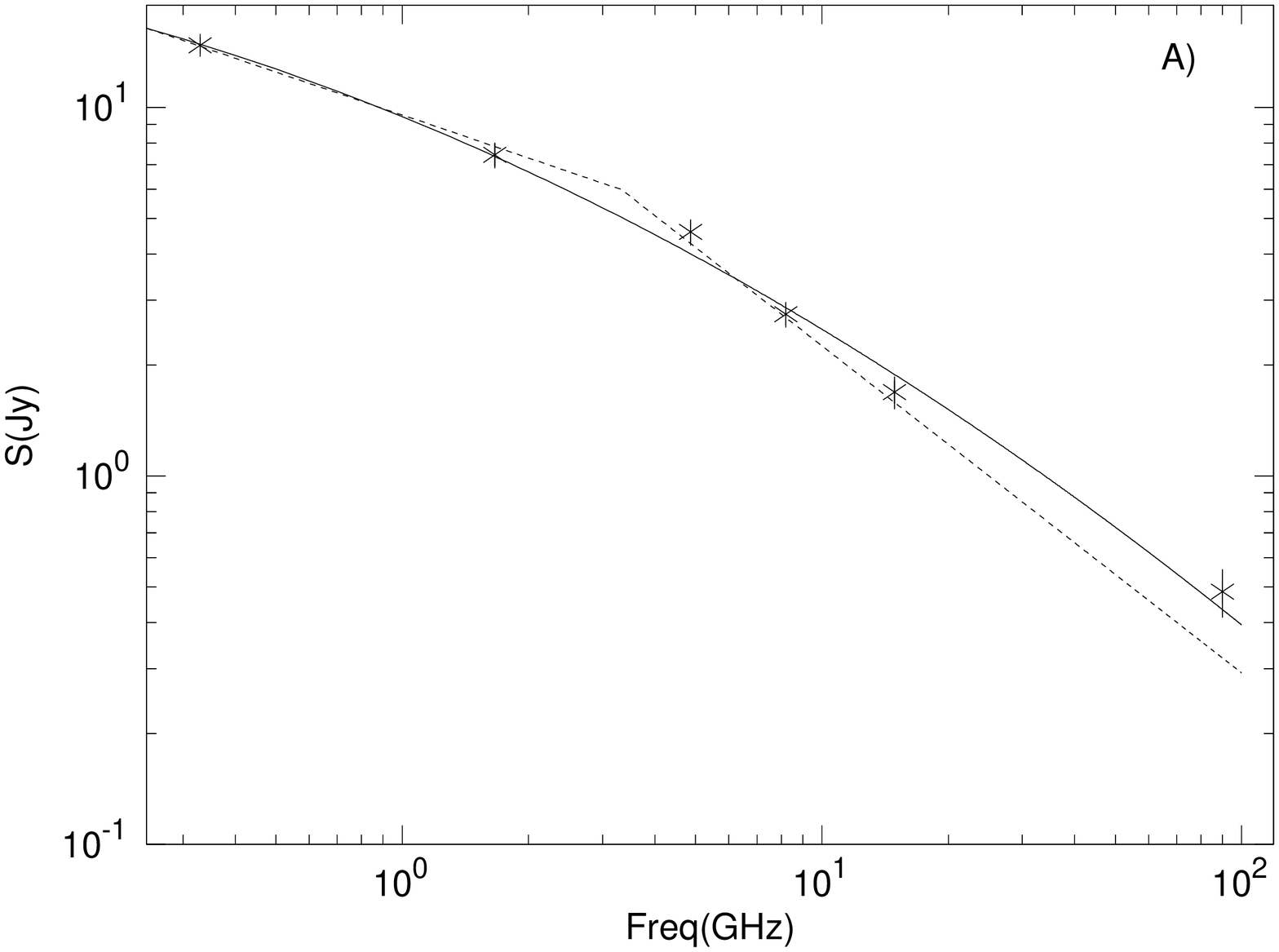}
\includegraphics[height=3.0in,angle=-90]{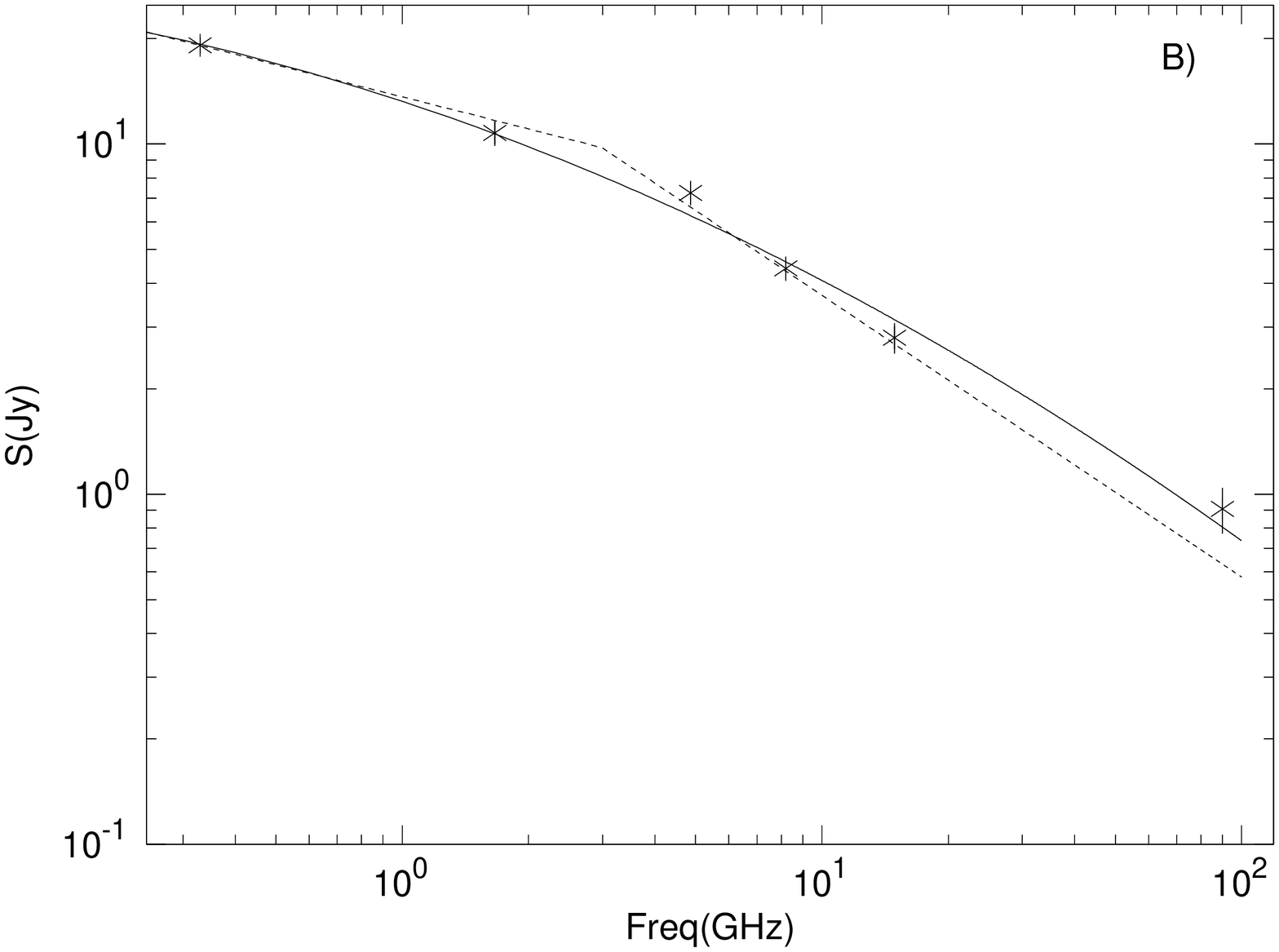}
}
\centerline{
\includegraphics[height=3.0in,angle=-90]{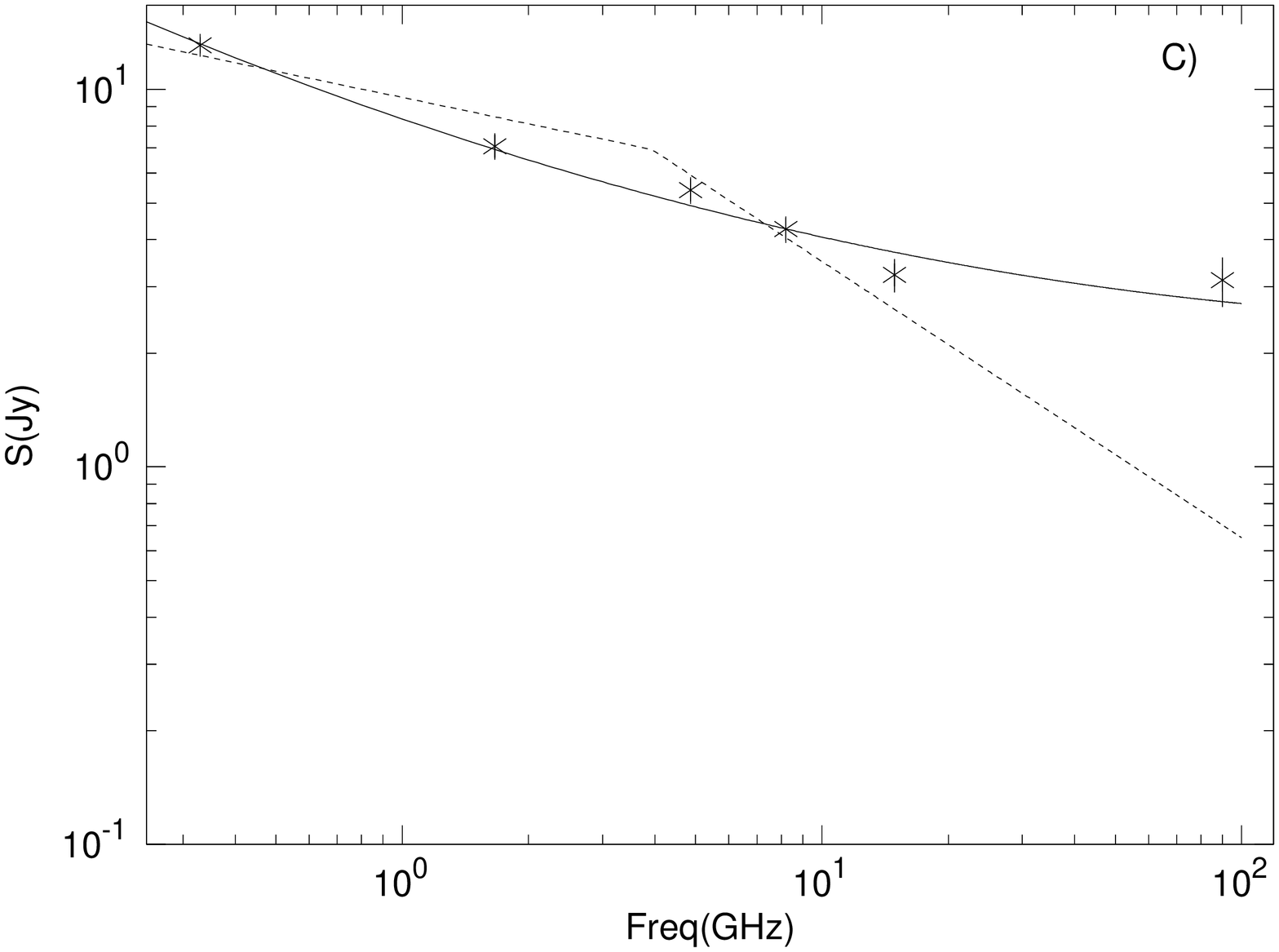}
\includegraphics[height=3.0in,angle=-90]{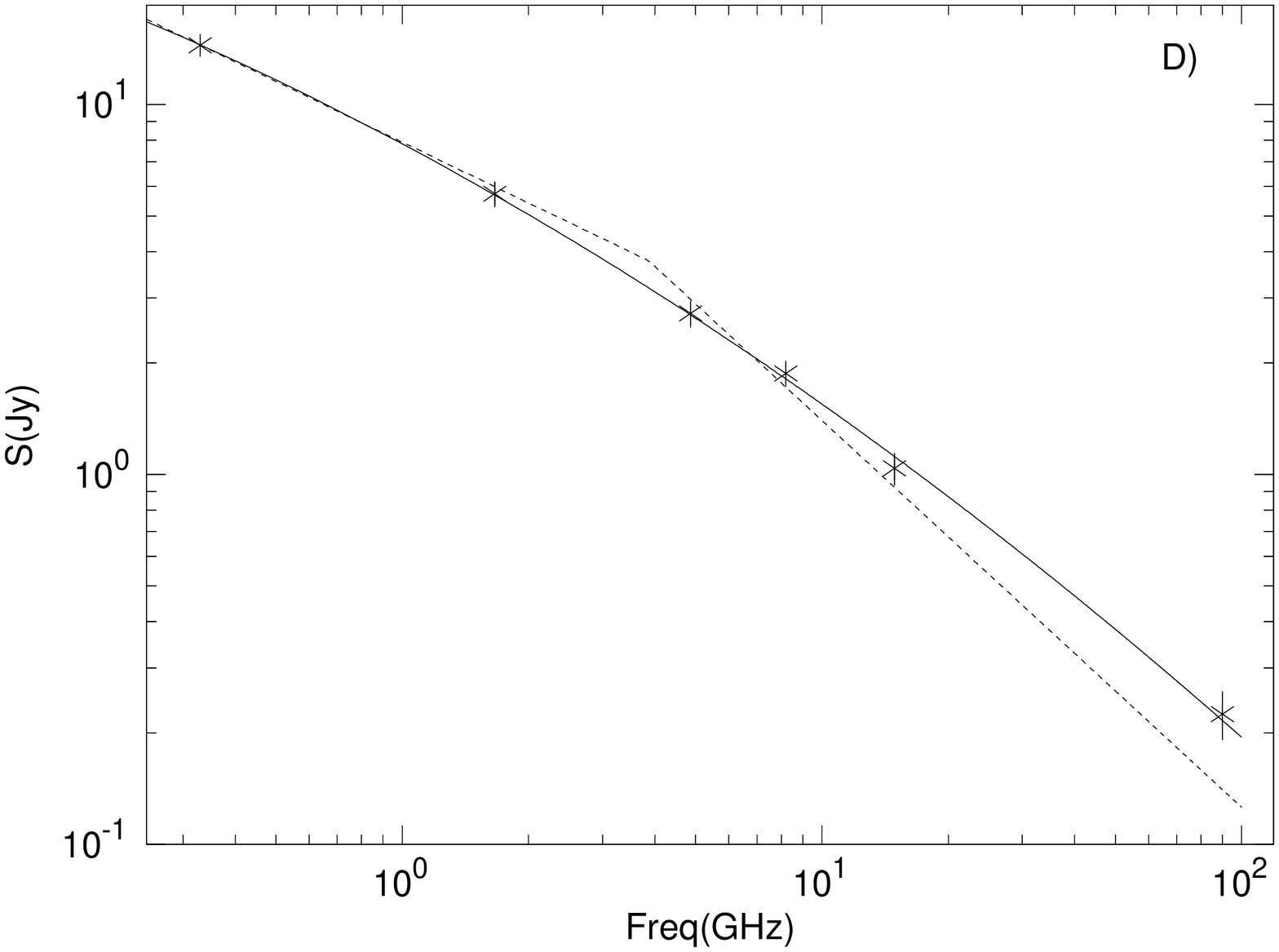}
}
\centerline{
\includegraphics[height=3.0in,angle=-90]{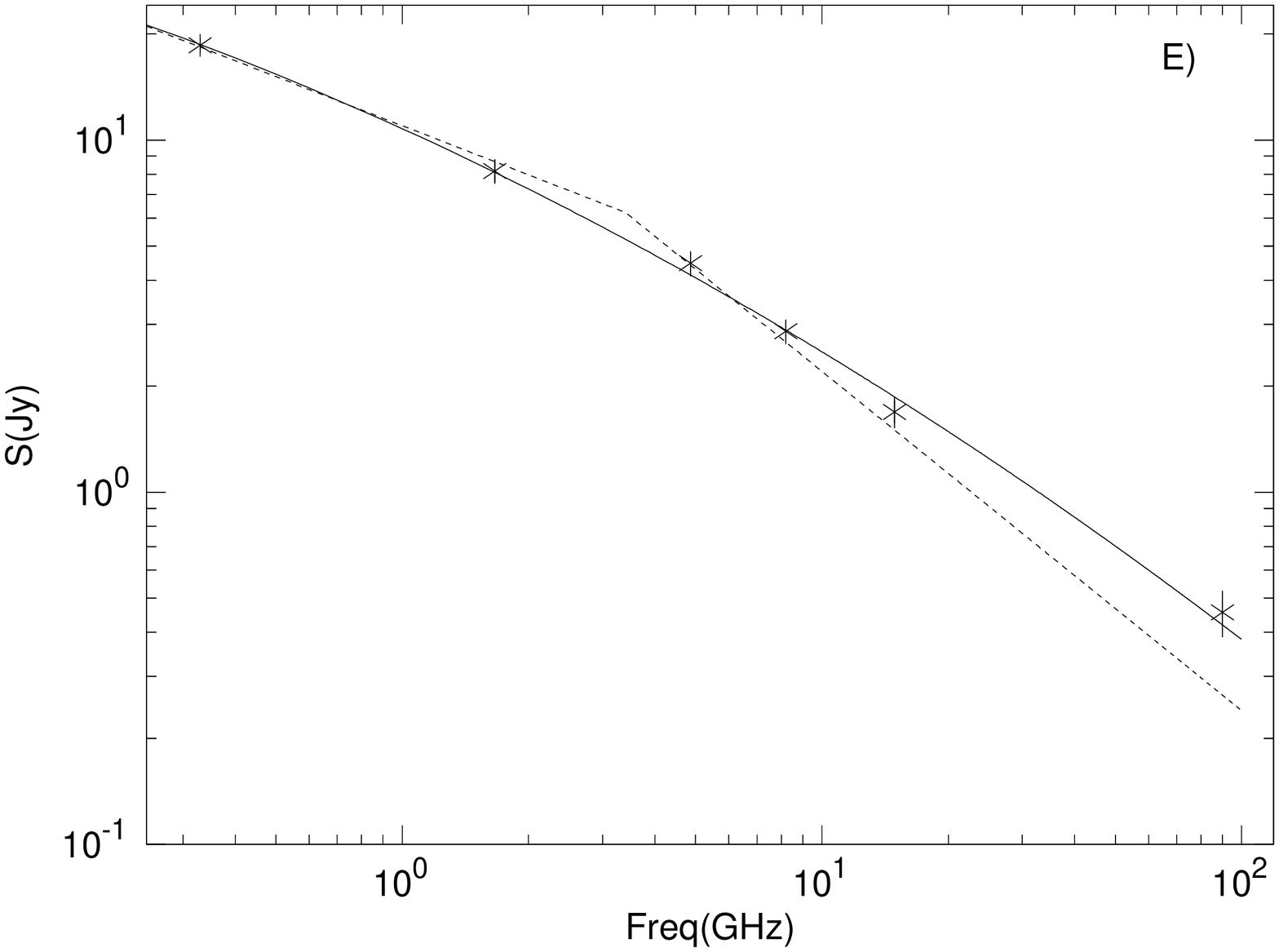}
\includegraphics[height=3.0in,angle=-90]{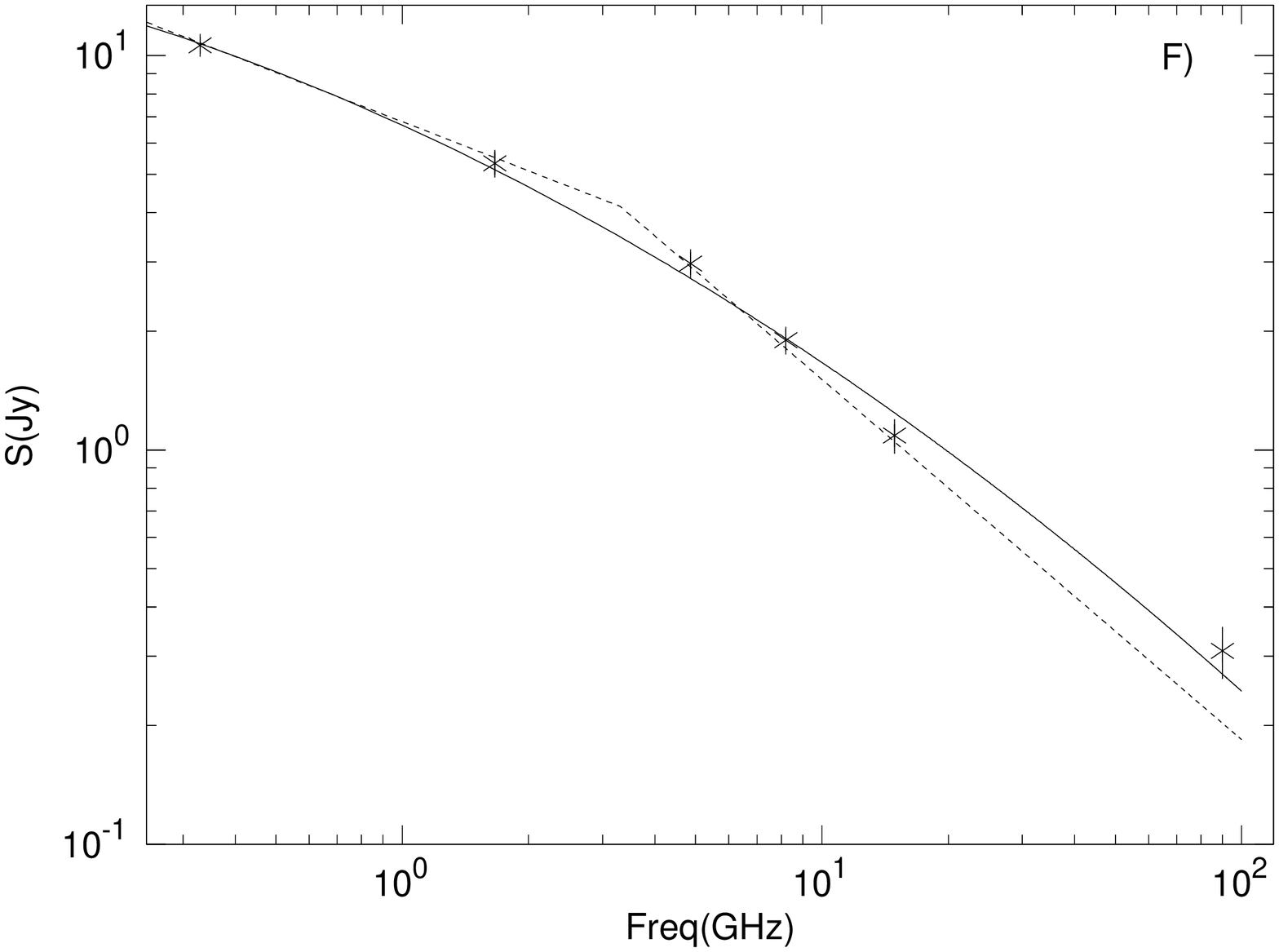}
}
\caption{ 
Point spectra of M87 at locations indicated in Figure \ref{M87ContFig}.
Fitted values are shown in Table \ref{tblM87Pt}.
Solid lines show the curved spectrum fits and the dashed lines
broken power law fits.
}
\label{M87PtSpectraFig}
\end{figure*}

\begin{figure}
\centering
\centerline{
\includegraphics[height=2.1in,angle=-90]{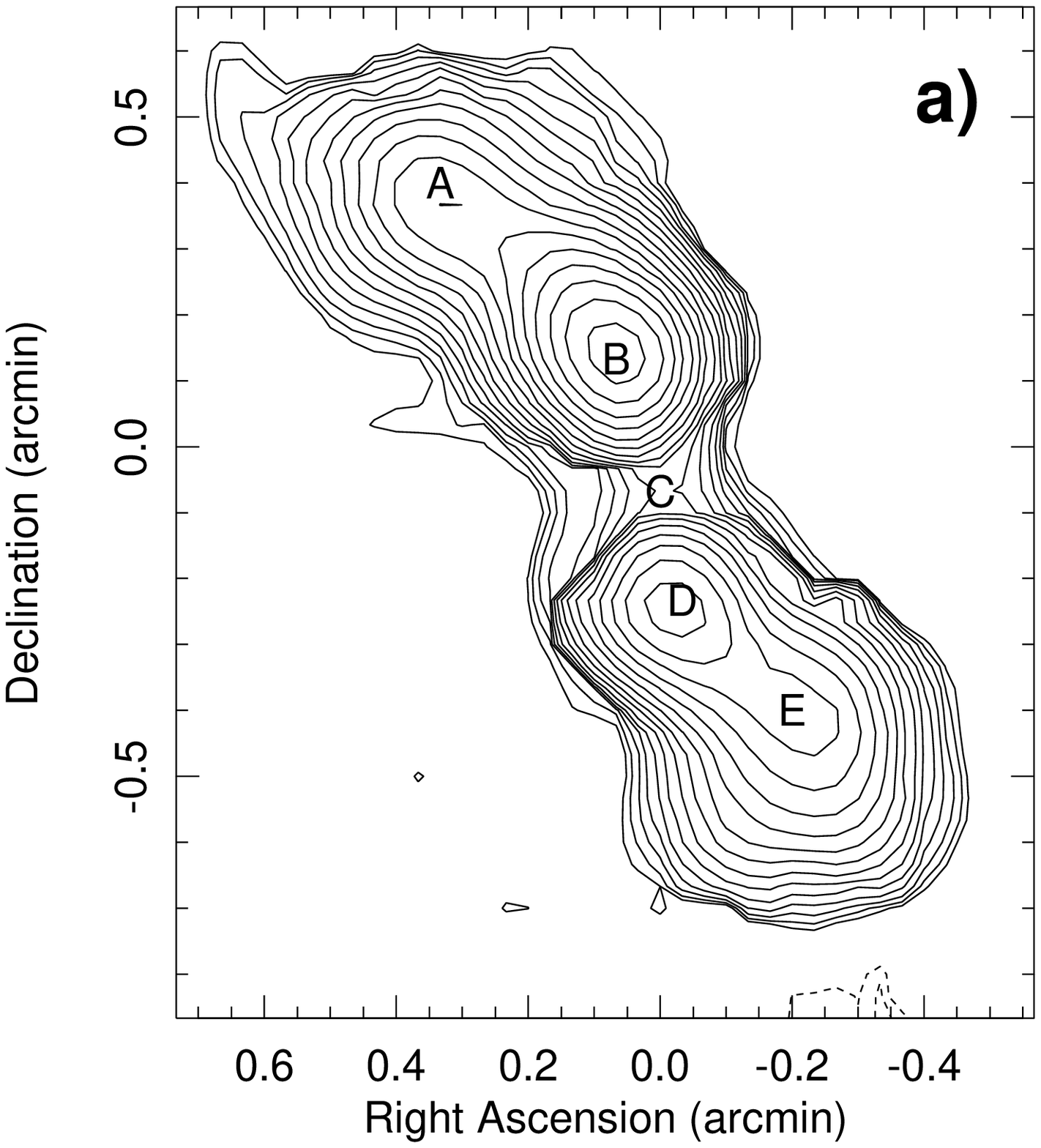}}
\centerline{
\includegraphics[height=2.4in]{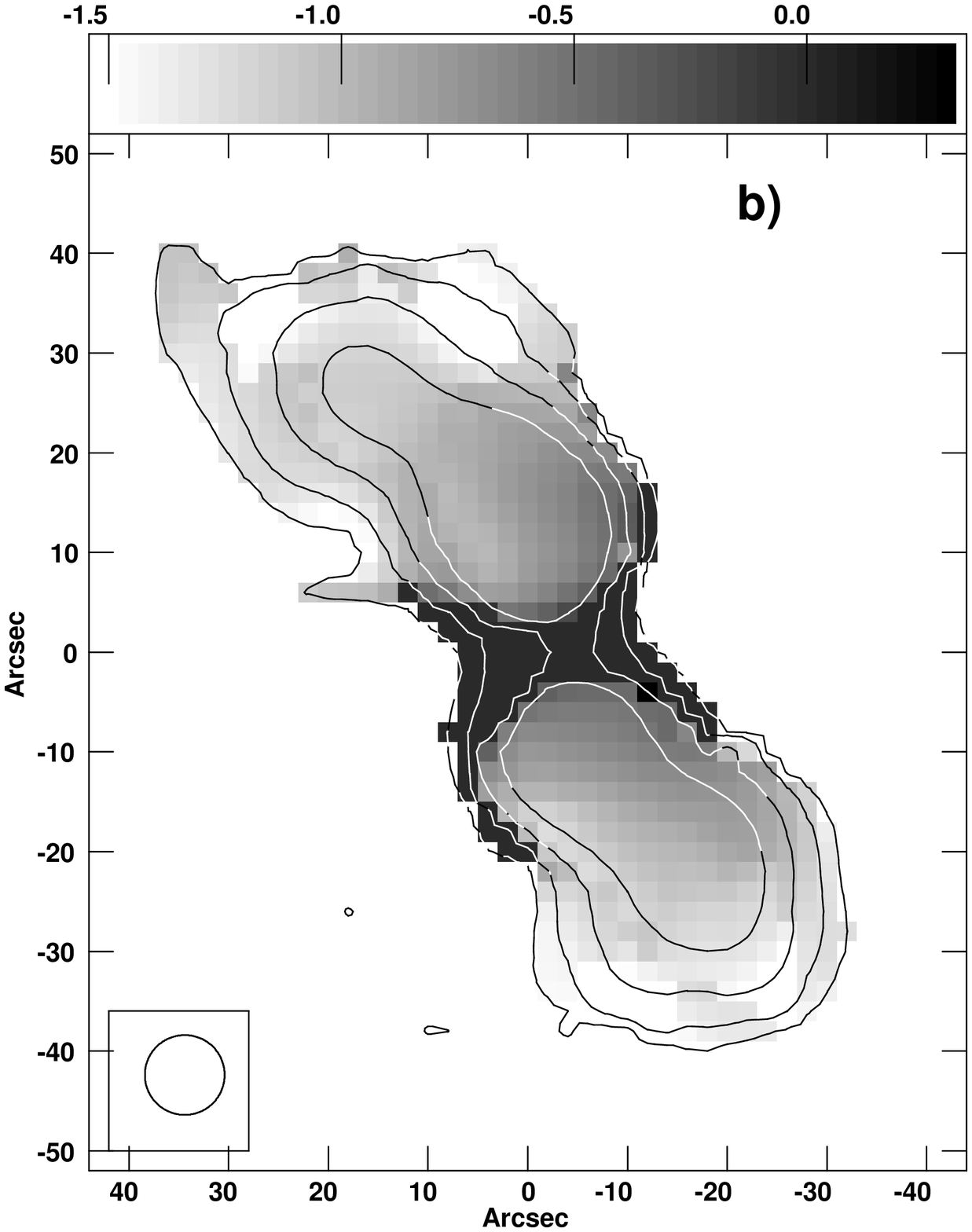}}
\centerline{
\includegraphics[height=2.4in]{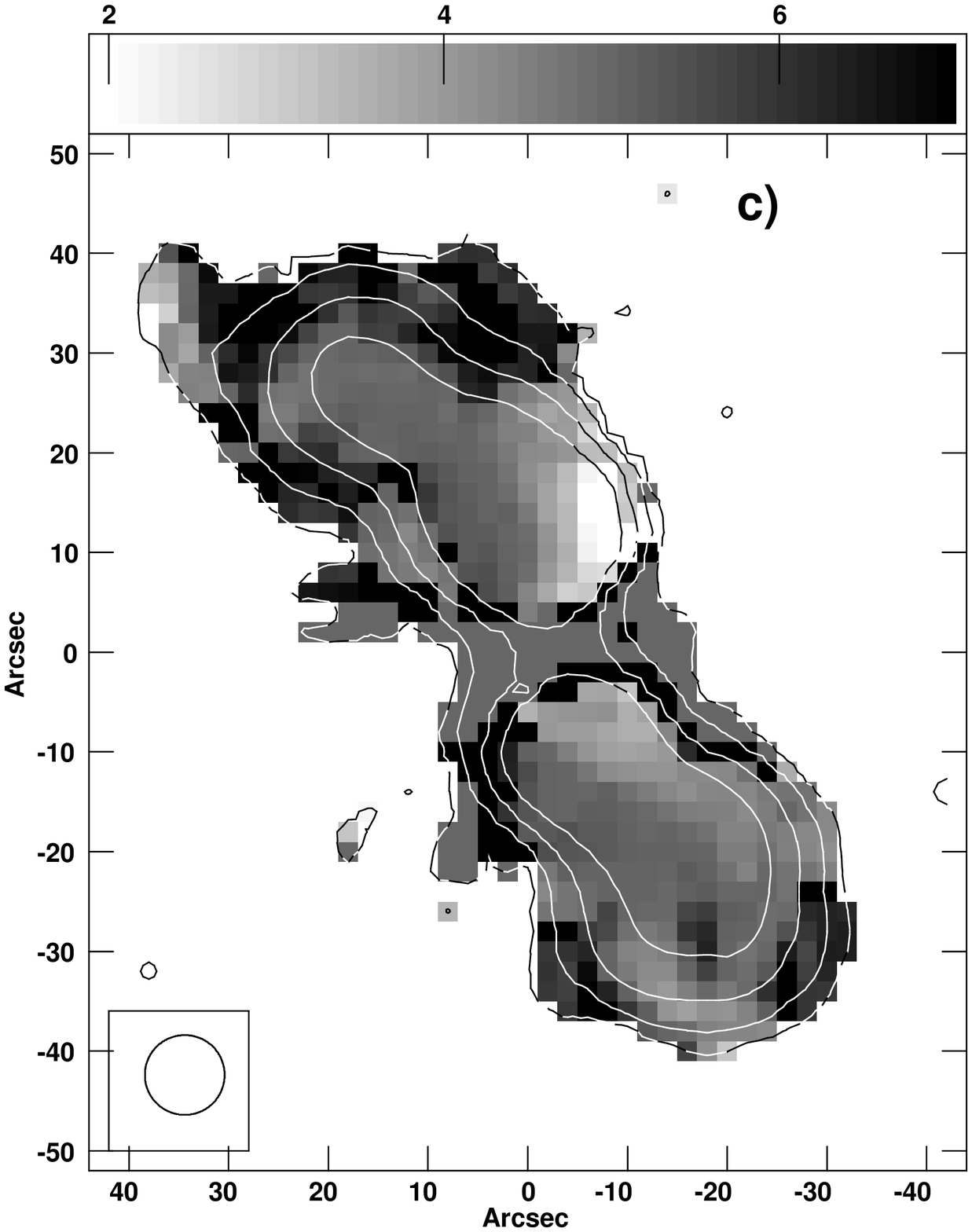}}
\caption{ 
{\bf a)} Fitted image of Hydra A at 5 GHz with indications of locations of point
spectra as letters.
Contours are powers of $\sqrt{2}$ times 10 mJy/beam.
{\bf b)} Grey-scale spectral index with contours of flux density at 10,
30, 100 and 300 mJy/beam over-plotted.
The gray-scale is given by the scale-bar at the top.
The resolution is indicated by the circle in the lower left corner.
{\bf c)} as {\bf b)} but break frequency (GHz) shown in gray-scale.
}
\label{HydraAContFig}
\end{figure}

\begin{figure*}
\centering
\centerline{
\includegraphics[height=3.0in,angle=-90]{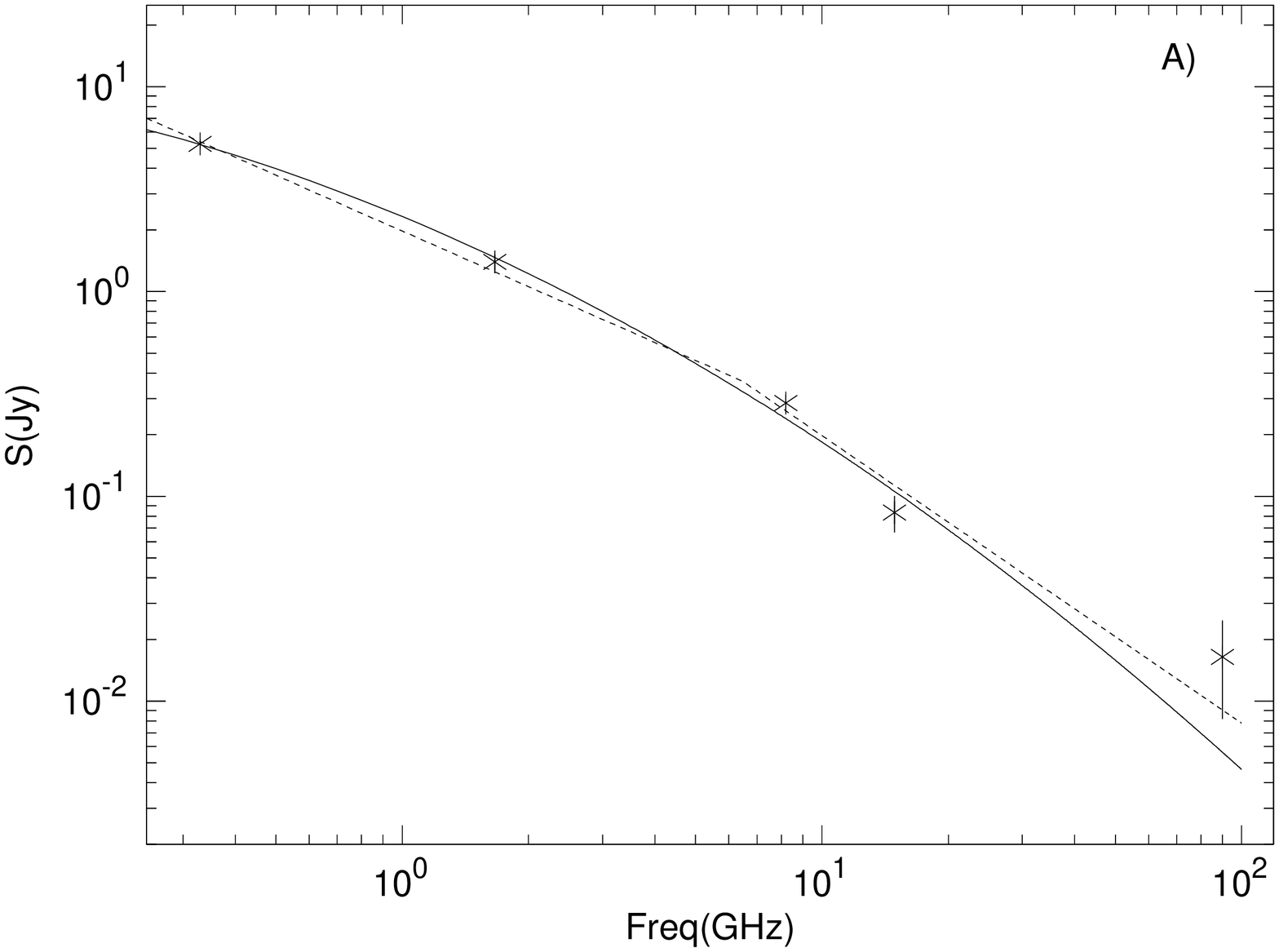}
\includegraphics[height=3.0in,angle=-90]{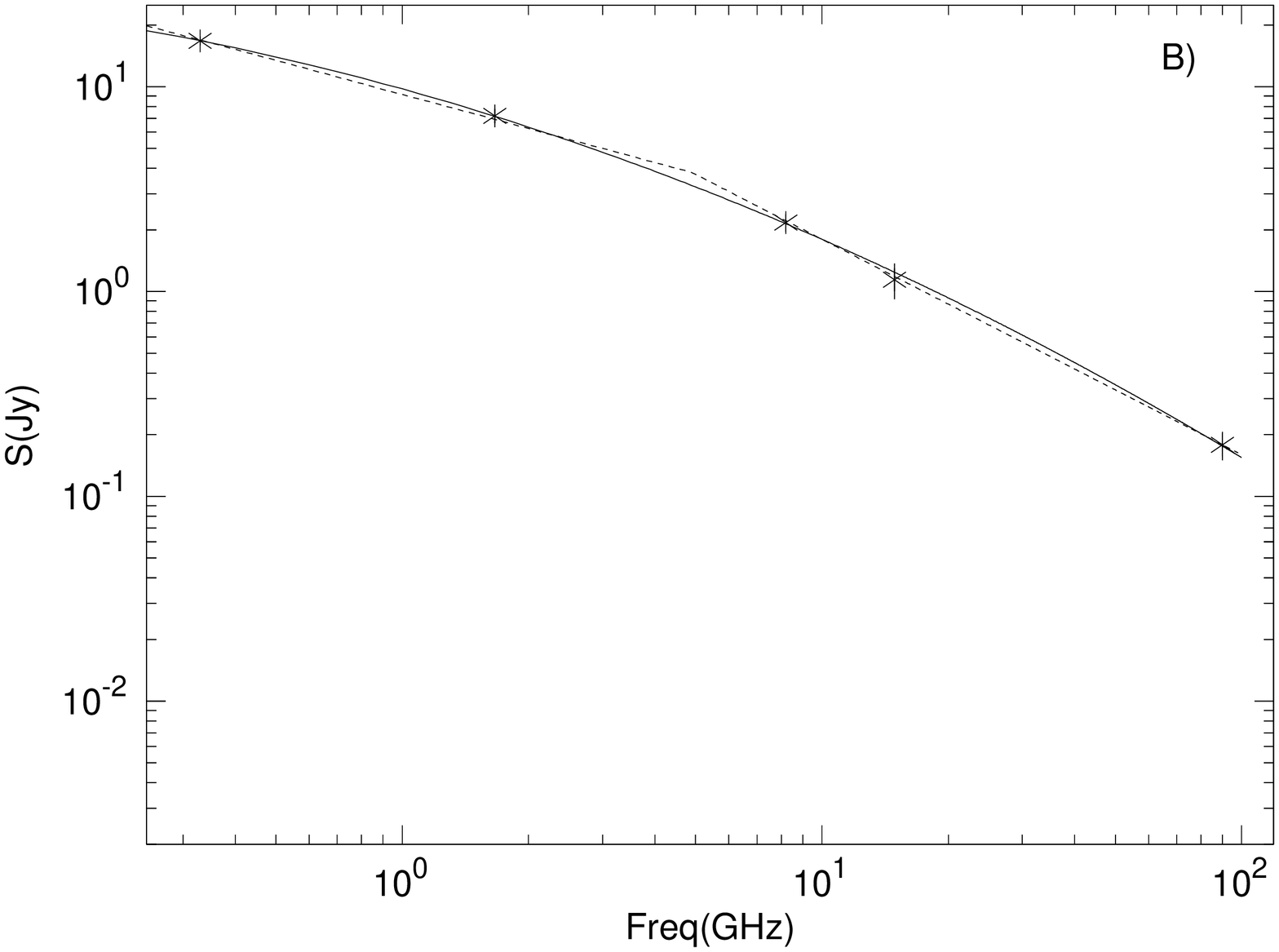}
}
\centerline{
\includegraphics[height=3.0in,angle=-90]{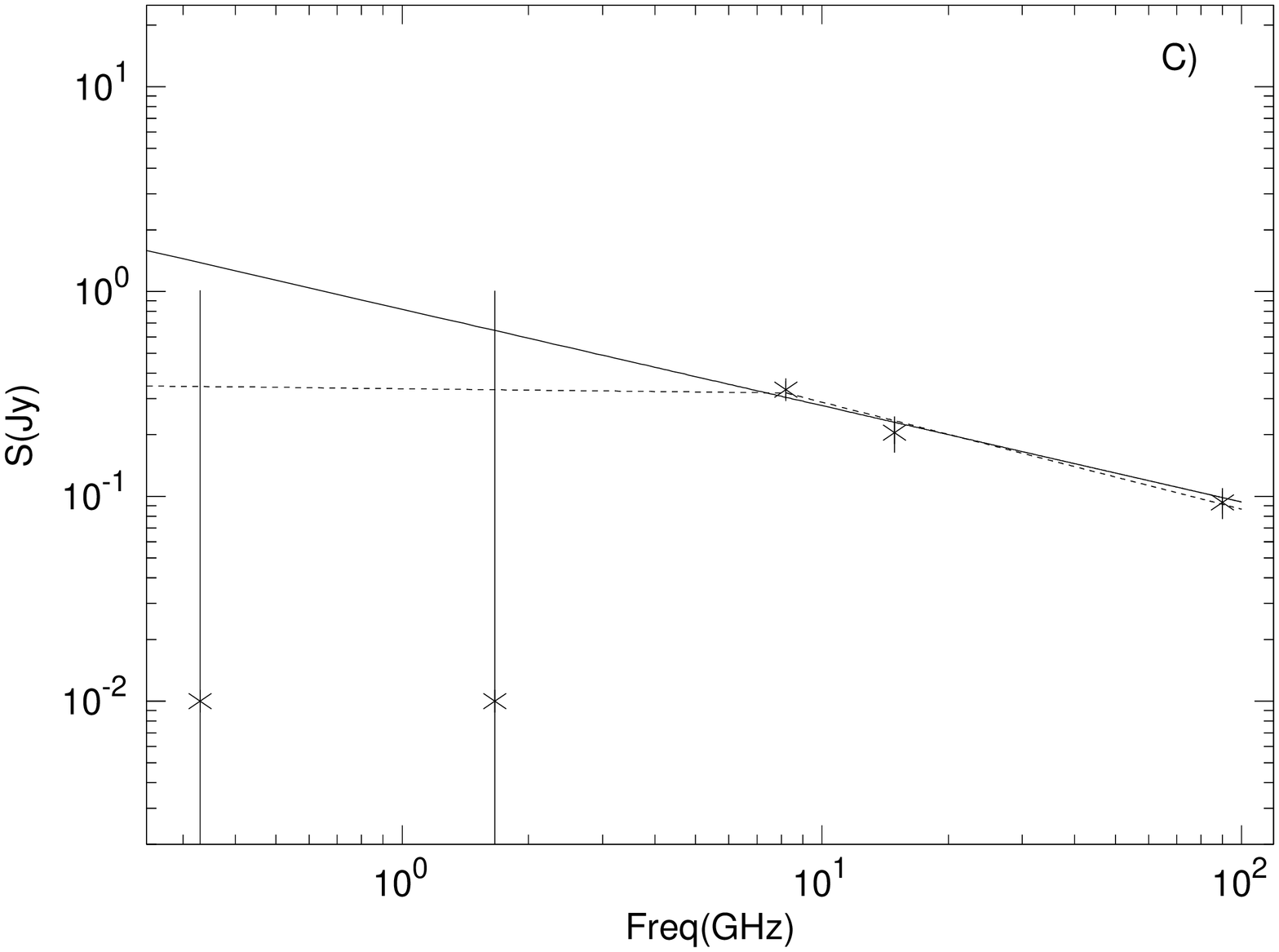}
\includegraphics[height=3.0in,angle=-90]{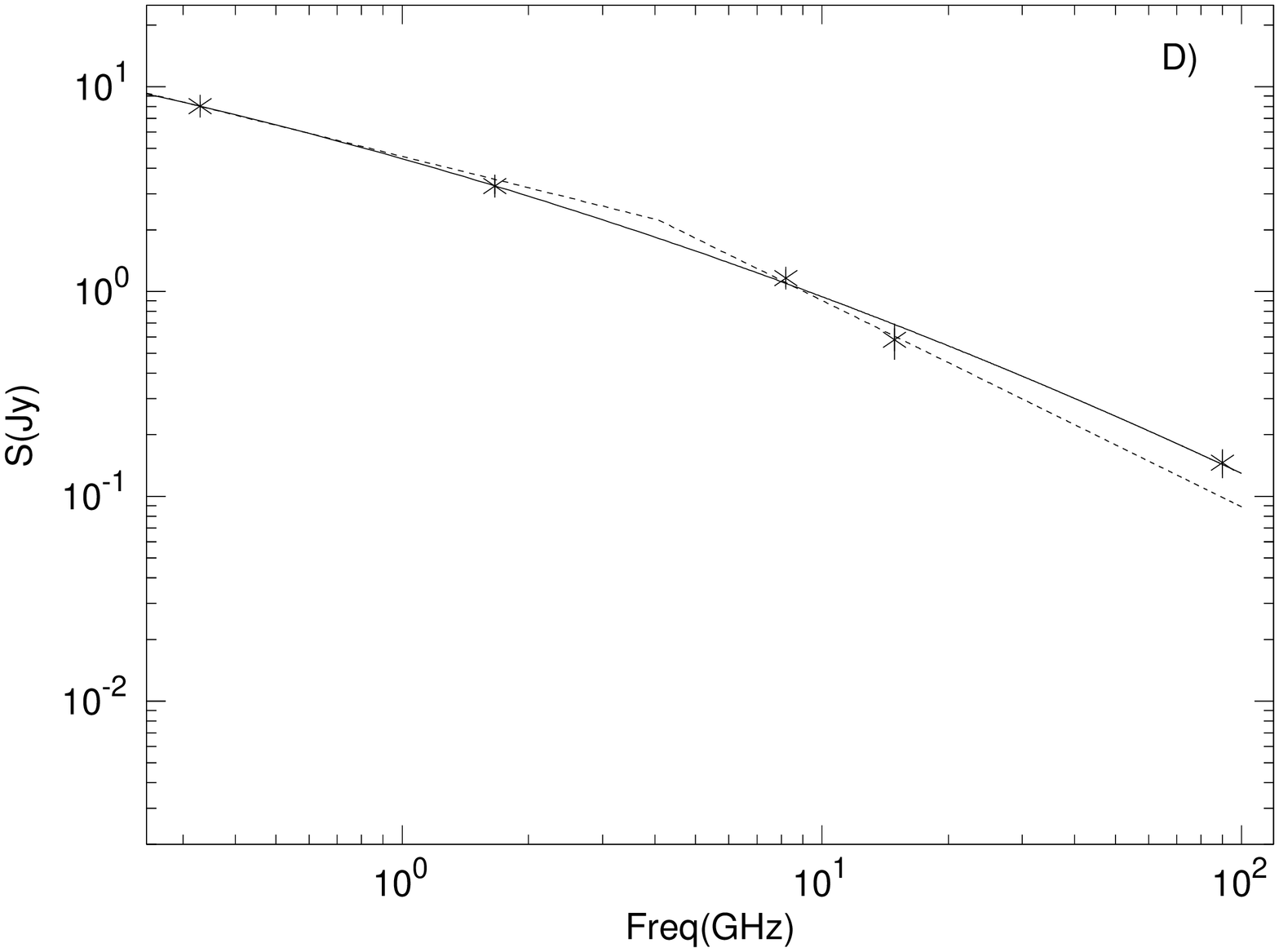}
}
\centerline{
\includegraphics[height=3.0in,angle=-90]{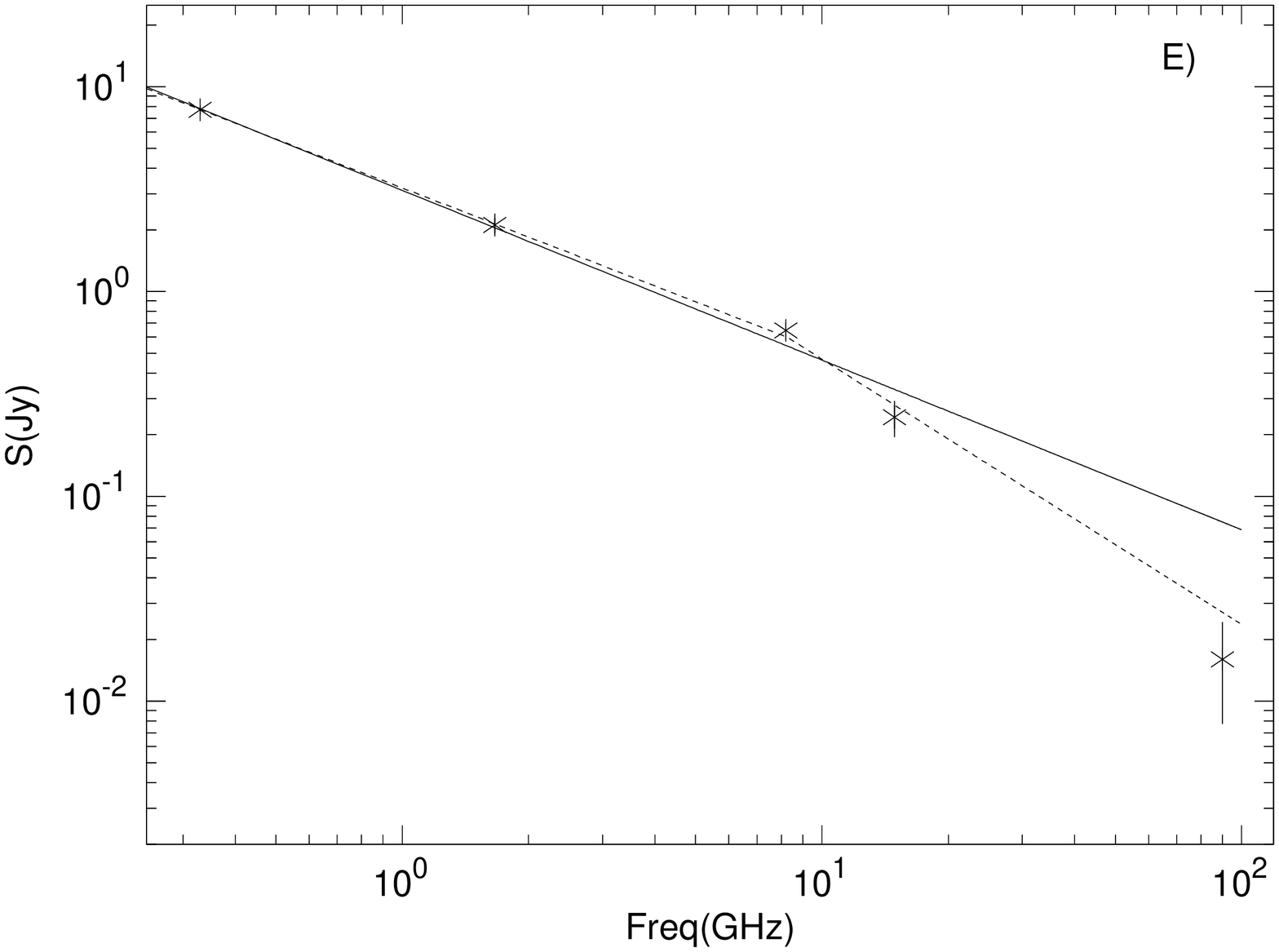}
}
\caption{ 
Point spectra of Hydra A at locations indicated in Figure \ref{HydraAContFig}.
Fitted values are shown in Table \ref{tblHydraAPt}.
Solid lines show the curved spectrum fits and the dashed lines
broken power law fits.
}
\label{HydraAPtSpectraFig} 
\end{figure*}

Plots of the derived image and spectral parameters are shown in Figures
\ref{M87ContFig} for M87 and \ref{HydraAContFig} for Hydra A.
Letters on Figures \ref{M87ContFig}a and \ref{HydraAContFig}a give the
locations where individual spectra were determined.
These individual point spectra are in Figures \ref{M87PtSpectraFig}
and \ref{HydraAPtSpectraFig} for M87 and Hydra A respectively.
Both continuously curved spectral fits and broken power laws are shown.
Corresponding fitted parameters are given in Tables \ref{tblM87Pt} and
\ref{tblHydraAPt}.
These tables give the flux density in Jy, spectral index and curvature
terms (when fitted), $\chi^2$ per degree of freedom of the curved
spectrum fit, the break frequency in GHz from the broken power law fit and
$\chi^2$ for the broken power law fits at the locations indicated in
Figures \ref{M87PtSpectraFig} and \ref{HydraAPtSpectraFig}.

\begin{table*}
\begin{center}
\caption{Hydra A Fitted Point Spectra at 5 GHz.\label{tblHydraAPt}}
\begin{tabular}{crrrrrr}
\tableline\tableline
Posn. & Flux & Spectral& Curvature & $\chi^2_{\rm curve}$ & Break Freq. & $\chi^2_{\rm break}$\\
 & density (Jy) &  index &  &  & GHz \\
\tableline
A & 0.4 $\pm$ 0.03 & -1.20 $\pm$ 0.08 & -0.11 $\pm$  0.03  & 4.7 & 6.4 $\pm$ 0.8 & 5.4 \\
B & 3.2 \phantom{$\pm$} 0.22 & -0.80 \phantom{$\pm$} 0.03 & -0.07 \phantom{$\pm$} 0.01 & 0.1 & 4.9 \phantom{$\pm$} 0.1 & 0.2 \\
C & 0.4 \phantom{$\pm$} 0.05 & -0.47 \phantom{$\pm$} 0.07 &                            & 1.1 & 8.2 \phantom{$\pm$} 3.2 & 0.4 \\
D & 1.6 \phantom{$\pm$} 0.11 & -0.71 \phantom{$\pm$} 0.03 & -0.04 \phantom{$\pm$} 0.01 & 0.6 & 4.1 \phantom{$\pm$} 0.1 & 2.7 \\
E & 0.8 \phantom{$\pm$} 0.05 & -0.83 \phantom{$\pm$} 0.03 &                            & 3.0 & 8.3 \phantom{$\pm$} 4.4 & 1.04 \\
\tableline
\end{tabular}
\end{center}
\end{table*}

\section{Discussion}
The presence of optical and X-ray emission from the inner jet in M87
\citep{Biretta91} necessitates a resupply of energetic particles in at least
this portion of the jet.
The data presented above can probe the possibility of further particle
resupply in the portions of the jet only detected in radio emission.

The radiative ``age'' of a source with break frequency of  $\nu_B$ is
given by \citet{Carilli91} as
$$ t\ =\ 1610B^{-1.5}\nu_B^{-0.5}\ \qquad\qquad {\rm 2)} $$
where B is in micro-gauss, $\nu_B$ in GHz and t in Myr.
Thus, the electrons radiating in a 10 micro-gauss field giving rise to
90 GHz radiation will start to be seriously depleted in about 5 Myr.

\subsection {M87}
One of the earliest cases recognized to require particle acceleration
outside of the nucleus was the optical jet in M87 \citep{Felten68}.
M87 is the dominant elliptical galaxy in the center of the nearby
Virgo cluster.
Previously published radio observations include \citet{Owen80},
\citet{Biretta83} and \citet{Owen00}.
We assume a distance to M87 of 16 Mpc which results in a linear
projected size of 78 pc per arc-second.
The jets are thus visible out to beyond 2 kpc from the nucleus in
Figure \ref{M87ContFig}.

M87 is an asymmetric two-sided FRI jet source in which both jets bend
into the line of sight limiting the distance from the nucleus to which
they can be observed.  Figure \ref{M87ContFig} contains the jets and
inner lobes embedded in the much larger region of emission shown by
\citet{Owen00}.  The jets on either side of the nucleus (position C)
show little variation in spectral index and curvature.  The
approaching jet (positions A, B) have somewhat flatter spectra than in
the receding jet (positions, D, E, F).  The fitted break frequencies
shown in Figures \ref{M87ContFig} and \ref{M87PtSpectraFig} and Table
\ref{tblM87Pt} show minimal variation along the visible portion of
the jets.  Furthermore, the continiously curved spectral model gives
consistently better fits, in terms of the $\chi^2$ per degree of
freedom than the broken power law spectra.

Magnetic field estimates based on minimum energy arguments and higher
resolution VLA images given by \citet{Hines89} for features in the
regions probed here, range from 50 to 100 micro-gauss.
This leads to estimates of radiative lifetimes of the 90 GHz emitting
electrons in the range 150 to 500 kyr.
We note that this is easily within the possible transport times
for the radiating particles from the nucleus.

The variation in spectral index and curvature across the source shown
in Figures \ref{M87ContFig} and \ref{M87PtSpectraFig} and Table
\ref{tblM87Pt} show very weak steepening with distance from the
nucleus. 
The curved nature of the spectra do indicate high energy losses to an
assumed initial power law electron energy spectrum.
The better fit of the curved spectrum than a simple broken power law
model  suggests a complex particle loss/resupply history rather than a
simple, radiatively evolving relativistic particle population.
Since the observable portion of the jet is relatively short, the lack
of evidence for electron aging could be due to either continued
particle acceleration along the jet or merely insufficient time for
the particle populations to evolve to the point they could be
detectable in our data.

The asymmetry of the spectral index is likely due simply to the 
emission on the counterjet side being stongly contaminated by the
steeper spectrum ``cocoon''.
The jet, being Doppler boosted, dominates the cocoon on its side.

\subsection {Hydra A}
The radio structure of Hydra A has been studied by \citet{Taylor90}
\citet{Dwarakanath95} and \citet{Lane04}; extended regions of very
steep spectrum emission are seen to the north and south of the
emission in Figure \ref{HydraAContFig}. 
This source is  a symmetric two-sided FRI radio galaxy with a weak,
strongly self absorbed core (position C). 
Hydra A is at a redshift of z=0.054 \citep{Dwarakanath95} giving a
projected linear size of 1.1 kpc per arc-second;
the source is visible in Figures \ref{HydraACleanFig} and
\ref{HydraAContFig} to approximately 45 kpc from the nucleus.

Magnetic field estimates of \citet{Taylor90} from minimum pressure
arguments range from about 10 to 70 micro-gauss in various features
giving estimated radiative lifetimes of the 90 GHz emitting electrons
in the range 0.3 to 5 Myr. 

In this galaxy, both the spectral index and curvature increase
dramatically along the visible portion of the jets, especially in the
north.  This suggests that particle acceleration slows or stops fairly
early in the jet and that the effects of particle aging are visible in
the outer parts of the source.  The magnetic field estimates by
\citet{Taylor90} suggest weaker field strengths in the northern
lobe. By Eq. 1, weaker fields should reduce the effect of electron
aging \citep[however, see][]{Blundell94}.  In the northern part of the
source (locations A and B) the curved and broken spectra are
comparable representations of the observed spectra whereas in the
south (location E), the broken power law spectrum more nearly
represents the high frequency measurements.  However, in spite of
strong variations in spectral index, there does not appear to be a
strong variation in the break frequency as represented in Figures
\ref{HydraAPtSpectraFig} and \ref{HydraAPtSpectraFig} and Table
\ref{tblHydraAPt}; if anything, the break frequency increases slightly
with distance from the nucleus.  This is more suggestive of a
decreasing magnetic field than an aging electron population.  The
apparent break in the spectrum in the south may be the result of
synchrotron losses in the inner regions of the jet where the magnetic
field is expected to be stronger.  This is inconsistent with a simple
intrepretation of the steepening along the jet being due to electron
losses; the fitted break frequency may not be meaningful except
possibly in the south where the broken power law is a better
representation of the spectrum than the curved model.  The curved
spectrum in the north suggests a complex electron history.

The observable portion of the jet in Hydra A is substantially longer
than in M87.
The longer time required to transport the radiating particles to the
outer regions allows radiative losses to accumulate.
This may be responsible for the apparent break in the spectrum in the
south.

\section{Conclusions}
We present new 90 GHz observations of the FRI AGN radio sources M87
and Hydra A at 8.5$''$ resolution and compare the results with lower
frequency archival VLA results at comparable resolution.
A spectral analysis is performed to look for evidence of particle
aging or re-acceleration.
M87 shows only weak variation out to 2 kpc but Hydra shows strong
steepening and bending over regions extending to 45 kpc.
This difference may be the result of the different size scales
observed in the two sources.
In M87, the length of jet observed may be too short for either the
particle resupply to damp out or for the radiating particles to age to
the point the the effects are visible in the observed spectral
region.
A continiously curved spectrum gives consistently better fits than a
broken power law spectrum.
This is more indicative of a complex history of the electron
energetics than a simple aging of a population of radiating
particles.

In Hydra A there is clear evidence of a depletion of higher energy
electrons, the loss mechanisms are winning over the resupply
mechanisms.  Even in the case of Hydra A, there is little evidence for
a variation in the break frequency with distance from the nucleus.
Only in the southernmost observed portion on the jet is there credible
evidence of a break in the spectrum.  The generally curved nature of
the observed spectra suggest complex histories of particle loss and
resupply.

\section*{Acknowledgment}
The National Radio Astronomy Observatory is operated by Associated
Universities, Inc. under cooperative agreement with the National
Science Foundation. This work was supported by a grant (AST-0607654)
from the National Science Foundation.

  \newpage



%

\acknowledgments




{\it Facilities:} \facility{GBT}.




\clearpage




\clearpage

\clearpage

\clearpage

\end{document}